\documentclass[12pt]{article}

\usepackage{amsfonts}

\def\hp{\hat{\pi}}
\def\hg{\hat{g}}

\def\N{{\cal{N}}}
\def\ep{\epsilon}
\def\E{\mathbb{E}}
\def\ve{\vskip0.7em}

\newcommand{\B}{\overrightarrow{B}}

\def\D{{\mathcal{D}}}
\def\cdot3{\cdot\cdot\cdot}

\def\half{\textstyle{\frac{1}{2}}}

\def\H{{\cal H}}

\def\ep{\epsilon}

\def\H{{\cal H}}
\def\B{\beta}
\def\E{{\rm I}\hskip-.2em{\rm E}}
\def\ra{\rightarrow}
\def\tint{{\textstyle\int}}
\def\hg{{\hat g}}
\def\hp{{\hat\pi}}

\def\s{\hskip.08em}
\def\dd{\partial}
\def\dag{\dagger}

\def\b{\begin{eqnarray*}}  %takes no eqn numbers
\def\e{\end{eqnarray*}}    %takes no eqn numbers
\def\bn{\begin{eqnarray}}  %takes eqn numbers
\def\en{\end{eqnarray}}   %takes eqn numbers

\def\<{\langle}
\def\>{\rangle}

\def\no{\nonumber}

\def\{{\lbrace}

\def\mathO{\mathcal{O}}
\def\om{\omega}
\begin{document} % NOW WITH ApP arxiv corrected

\title{ A Straight Forward Path to a  \\
Path Integration of Einstein's Gravity}

\author{John R. Klauder\\
Department of Physics and Department of Mathematics  \\ 
University of Florida,  Gainesville, FL 32611-8440}
\date{ }

\bibliographystyle{unsrt}

\maketitle

\begin{abstract}
%{\bf Forward:} 
Path integration is a respected  form of quantization that all theoretical quantum physicists should  welcome. This elaboration begins with simple examples of three different versions of path integration. After an important clarification of how gravity can be properly quantized, an appropriate path integral, that also incorporates necessary constraint issues, becomes a proper path integral for gravity that can effectively be obtained. How to evaluate such path integrals is another aspect, but most likely best done by computational efforts including Monte Carlo-like procedures.

\end{abstract}

\section{Introduction}
This article is an updated review of path integral procedures that can deal with quantizing gravity. However, it is useful to start with much simpler examples to learn just what will be needed to tackle gravity. This prelude contains what canonical quantization (CQ) and affine quantization (AQ) can (and cannot) do for toy models, which are then carried over to a valid path integration of gravity.\footnote{A summary of soluble path integrations, at least until 1993, is given in [1].}

For clarity, let us first begin with some  background regarding integration and quantization.

\subsection{The task of integration}
The standard mathematical integration symbol, $A=\tint_a^b f(x)\;dx$, may be easy or difficult to evaluate  depending on $f(x)$. For continuous functions, if you know that $f(x) = dg(x)/dx$, then the answer is $g(b)-g(a)$. Or there may be a simple trick like dealing with $B=\tint_{-\infty}^\infty e^{-x^2}\,dx$ by squaring it and passing over to polar coordinates. But the overwhelming bulk of integrations require expressions like $C=\lim_{N\ra\infty} \Sigma_{n=1}^N
[f(n\ep) +\mathO(\ep) ] \;\ep$, with $ N\ep$ fixed, which, at best, may be adequately evaluated using a computer which can reach large  $N$, but, unfortunately, not $N=\infty$.\footnote{An important article that emphasizes proper quantization 
is [2].}

Some of our final results will, of necessity, involve traditional integration-like, i.e., summation expressions.

\subsection{Full, and reduced, coordinate space for \\canonical quantization}
\subsubsection{A full coordinate space, $-\infty<q<\infty$}
Canonical quantization (CQ) requires classical variables $-\infty <p\;\&\;q<\infty$, with a Poisson  bracket  $\{q,p\}=1$,\footnote{A Poisson bracket of $A(p,q)$ and $B(p,q)$ is defined by $\{A(p,q), B(p,q)\} =\dd A(p,q)/\dd q\;\dd B(p,q)/\dd p -\dd A(p,q)/\dd p\;\dd B(p,q)/\dd q$. For our use it establishes the `size' of $q$ relative to that of $p$. For example, if  $q'=aq$, and $p'=bp$, then $\{q',p'\}=ab$. For our purpose, we then choose $ab=1$.} to
promote to 
quantum operators, $p\ra P$ and $q\ra Q$, which obey $[Q, P]=i\hbar1\!\!1$. While this can lead to quantum operators, only classical variables that are Cartesian, e.g., $d\sigma^2= \om^{-1}\,dp^2+\om\,dq^2$ are required [3] to lead to a valid quantization in which the quantum 
Hamiltonian  ${\cal{H}}(P,Q)$ has a functional equality with the classical Hamiltonian, i.e., $H(p,q)={\cal{H}}(p,q)$, when $\hbar\ra0$.

A familiar example of these rules is the harmonic oscillator, where, with unit coefficient factors, i.e., $m=\omega=1$, the classical Hamiltonian is $H=(p^2+q^2)/2$, and the quantum Hamiltonian is ${\cal{H}}=(P^2+Q^2)/2$, with eigenvalues given by $E_n=\hbar(n+1/2)$, where $n=0,1,2,...$, and eigenfunctions that are given by $\psi_n(x)=N_n \,e^{x^2/2\hbar} \,(-d/dx)^n e^{-x^2/\hbar}$, 
with $N_n$ providing normalization, 
i.e. $\tint_{-\infty}^{\infty}  |\psi_n(x)|^2\;dx =1$.  These eigenfunctions are even functions, i.e., $\psi_{2n}(-x)=\psi_{2n}(x)$, and odd functions, i.e., $\psi_{2n+1}(-x)=-\psi_{2n+1}(x)$, where in each case $n=0,1,2,...$. These even and odd eigenfunctions will be mentioned in a later section.

\subsubsection{A reduced coordinate space,  $0<q<\infty$}
This case is the half-harmonic oscillator with $q>0$, while the classical Hamiltonian is still $H=(p^2+q^2)/2$. The motion is like that of a half-pendulum in which, now like a tennis ball bouncing off a wall, 
our half-harmonic oscillator rebounds from a wall where $q=0$, and, at that point, $p$ changes direction. 

Quantization by CQ for this model fails, as we now demonstrate. If we choose CQ, you must {\it crush with a virtual wall} the wave function  to be  {\it zero for all}   $x<0$. Only the former positive half of an  `odd' function  (loosely named here, but named for when it was part of the whole real line) appears as a possible eigenfunction
 of the half-harmonic oscillator 
since it is continuously connected to the squashed wave function. However, when the second $P$ acts on the `new' wave function, now named an `even' function, there is a real gap in this wave function values at  $\psi'(0)$  since, for $x<0$ it is zero, while  if $x>0$, then $\psi'_{\lim_{x\ra0}}(x)=c \neq0$, which ensures that the `new' wave function is not continuous. The second $P$ differentiation then leads to a Dirac delta   $\delta(0)=\infty$, and thus the second derivative of the wave function becomes non-normalizable, and therefore is not allowed in any  Hilbert space, all of which implies a failure of canonical quantization for the half-harmonic oscillator using self-adjoint canonical operators.

If instead, we choose to completely  ignore the portion where $x<0$, and keep only $x>0$, then the operator $P^\dag\neq P$ which means there are infinitely many distinct quantum, self-adjoint, Hamiltonian operators each of which would pass to the same quantum Hamiltonian if only
  $P^\dag = P$.\footnote{An example of one set of infinitely many self-adjoint quantum operators that would reduce to $(P^2 + Q^2)/2$ if $P^\dag= P$, but instead since $P^\dag\neq P$, it 
  is given by $[(P^{\dag\,4+n}/P^{2+n}+P^{4+n}/P^{\dag \,2+n})/2+Q^2]/2$, with $n=0,1,2,...$.}
  
The integral $\tint^A_B (d/dx) \,[ f(x) g(x)] \;dx=f(A)g(A)-f(B)g(B)$. To eliminate that answer we choose $g(A)=g(B)=0$, but that leaves $f(A)$ and $f(B)$ free; this is like  $P^\dag$. To ensure that $P^\dag=P$, we then insist that $f(A)=f(B)=0$ as well.   
  
\subsection{Reduced coordinate space for affine quantization}
\subsubsection{A reduced coordinate space, $0<q<\infty$}
We start by accepting that $P^\dag \neq P$, and
 first choosing a substitute for the classical variable $p$
 which leads to the dilation variable, $d=pq$, which first requires that  $q\neq0$,  because if $q=0$ then $d=0$ and
$p$ can not help. For our model, we then discard $q<0$ and keep only $q>0$. It follows that $d=pq\ra D=(P^\dag Q+QP)/2=D^\dag$, along with $Q=Q^\dag>0$. In addition we find that  $[Q,D]=i\hbar Q$, a Lie algebra that is formally  similar to that of the affine group, and which we have accepted as the same name for our quantum procedures.
Just as analysis has confirmed that $H(p,q)\ra H(P,Q)$ for CQ, it has also determined that $H'(d,q)\ra H'(D,Q)$ for AQ, i.e., for an affine quantization [4]. 

 For our half-harmonic oscillator model we choose the classical Hamiltonian as $H' =(d^2/q^2 +q^2)/2$, and $q>0$, which is then quantized to become 
   \bn {\cal{H}}'=(DQ^{-2}D +Q^2)/2=[P^2+(3/4)\hbar^2/Q^2+Q^2]/2 \;,\en where
   the `3/4' term also implies that $P^\dag$ and  $(P^\dag)^2 $ {\it act} like $P$ and $P^2$ as needed in this equation. The eigenfunctions of this equation have eigenvalues that are $E'_n =2\hbar(n+1)$, with $n=0,1,2,...$
   [5]. This behavior leans toward the full-harmonic oscillator eigenvalues and eigenfunctions if the space  of $x>0$ is extended 
to $x>-b$, where $b>0$, and then the quantum Hamiltonian becomes $
{\cal{H}}'_b=[P^2+(3/4)\hbar^2/(Q+b)^2+Q^2]/2$. Finally, we can let $b\ra \infty$, which reaches the full-harmonic oscillator, with  all of the original eigenvalues and eigenfunctions being fully recovered 
[6].\footnote{In this authors's article [6], page 15, Fig.~1, there is an elegant graph that captivates the set of equally spaced eigenvalues, which are plotted for different $b$-values.} 

Recovering the full-harmonic oscillator when $b\ra\infty$ would not have been possible when using CQ.

\subsection{The relevance of the previous section  to \\quantum gravity}
The reader may be wondering what the half-harmonic oscillator has to do with quantum gravity. The answer is that the metric field,  $g_{ab}(x)$, is physically positive, such as $ds(x)^2= g_{ab}(x)\;dx^a\;dx^b>0$, provided that $\Sigma_a (dx^a)^2>0$. The 
metric field can be diagonalized by nonphysical, orthogonal matrices 
and their transpose ($T$) matrices -- using matrix notation, i.e., $G(x)=\{g_{ab}(x)\}$, this means that $O(x) \,G(x) \,O(x)^T =G_d(x)$  -- which leads to a diagonal metric ($G_d(x)$) that requires each of the three diagonal metric terms to be strictly  positive, a requirement 
not unlike that of the half-harmonic oscillator's requirement that $q>0$. 

It may seem insignificant to change from $q>0$ to  $ q\geq0$ , or from $g_{ab}(x)>0$ to $g_{ab}(x)\geq0$, but that is not the case. Ask yourself: Is it true that $q\times q^{-1}= 1$  for {\it all} $q$ including $q=0$, or $g^{cb}(x)g_{ab}(x)= \delta^c_a$ for {\it all} $g_{ab}(x)$ including $g_{ab}(x)=0$?  To be mathematically safe, and physically correct, it is necessary to accept  $q>0$ as well as  $g_{ab}(x)>0$.\footnote{To accept that $g_{ab}(x)\geq0$  shows the `gravity of the issue'.}

\section{An Overview of Path Integration \\Procedures}
In a certain sense there are three, somewhat different, avenues to choose when considering doing a path integration. 

The first approach, in Sec.~2.1, is a completely formal attempt to provide a result that may, or may not, be absolutely correct. This procedure is the simplest of all but its formal status and its lack of precision needs to be considered. In Sec.~2.1 we offer a brief examination of these formal efforts.

The second approach, presented in Sec.~2.2, is one in which the path integration procedures are considered in a non-regularized, and using completely precise procedures, are offered in the form of functional integrations that 
effectively deal with the problem fully from a mathematical perspective. However, precise mathematics need not guarantee precise physics.  

Our third, and last approach, which we will develop fully later in Sec.~3.2.4, is to use mathematical-physics aspects in a clear and natural approach which is guaranteed to offer proper and valid path integration results. 

All of this analysis is to prepare us for a path integration of gravity.

\subsection{Canonical path integrations}
The original path integration expression for a canonical quantization of a typical classical action is given, with $-\infty<p\;\&\;q<\infty$, by
   \bn \<q'';T|q';0\>= \N\int e^{(i/\hbar)\tint _0^T\{ p(t)\dot{q}(t) -[ p(t)^2/2+V(q(t))]\,\}\:dt}\;\D(p)\;\D(q)\;, \en
   where $x= q$ and $|x\>$ is a ket for the Schr\"odinger representation 
   where $Q|x\>=x|x\>$. The integrations cover very general  functions that pass from $q'=q(0)$ to $q''=q(T)$. These paths include integrable infinities, such as the case where $p(t) = A/|t-1|^{1/3}$, while  there can be  cases where $\dot{q}(t)= B/|t-2|^{3/4}$, etc. Both of these examples are forbidden by Wiener measures!
   
   Although there are certain features that might complicate this formulation of path integration, there also are rigid mathematical formulations as well such as that in the next section.
  \subsection{Introduction of Wiener-type measures}
An example of a mathematically acceptable canonical path integral includes two (formally expressed) Wiener measures [7 - 9]. Again, for a canonical quantization, the new formulation is given by
 \bn &&\<q'';T|q';0\>_{CQ}=\lim_{\nu\ra \infty}\N_\nu\int e^{(i/\hbar)\tint _0^T\{ p(t)\dot{q}(t) -[p(t)^2/2+ V(q(t))]\,\}\:dt}\no \\
 &&\hskip5em \times e^{-(1/2\nu)\tint_0^T [\omega^{-1}\,\dot{p}(t)^2 +\omega\,\dot{q}(t)^2]\;dt}\;\D(p)\;\D(q)\;. \en
 Here, the Wiener measures serve to better control the family of paths that cover the overall path integral. We can assume that $\omega$ has the dimensions of $\hbar$, while $\nu$ has the dimension of $\hbar/time$. Indeed, the Wiener measure enables us to link the classical and quantum expressions quite well mathematically.

 Wiener-like measures can also be used to provide mathematically sound affine path integrations as well, and where $q(t)>0$, such as
  \bn &&\<q'';T|q';0\>_{AQ}=\lim_{\nu\ra \infty}\N'_\nu\int e^{(i/\hbar)\tint_0^T\{ p(t)\dot{q}(t) -[ p(t)^2/2+V(q(t))]\,\}\:dt}\no \\
 &&\hskip5em \times e^{-(1/2\nu\hbar)\tint_0^T [\B^{-1}\,q(t)^2\,\dot{p}(t)^2 + \B \,q(t)^{-2}\,\dot{q}(t)^2]\;dt}\;\D(p)\;\D(q)\;, \en
 in which we have traded the Wiener flat metric for a Wiener-like 
 constant negative curvature surface measure. In this equation, $\B$ can have the dimension of $\hbar$ and $\nu$ has the dimension of inverse time.
 
 \subsubsection{An affine path integral of the half-harmonic oscillator}
 The classical Hamiltonian of the half-harmonic oscillator
  is, as Eq.~(1) notes, still $H=(p^2+q^2)/2$, but the important property now is that $q>0$ and any path $q(t)$ just `bounces back off a `virtual wall' $q=0$. The affine path integration of this model is given by\footnote{An alternative route to a novel path integration of this example chooses a special `canonical route' that includes a semi-classical Hamiltonian,  $H =[p^2+(3/4)\hbar^2/q^2+q^2]/2$, in which a canonical-like path integral, with  $q>0$,
 is allowed, thanks to the necessary `3/4' semi-classical term [5, 6].}
   \bn &&\<q'';T|q';0\> =\lim_{\nu\ra \infty} \N_\nu\int e^{(i/\hbar)\tint_0^T[p(t)\dot{q}(t)-(p(t)^2+q(t)^2)/2 ]\;dt}\no \\
   &&\hskip4em \times e^{-(1/2\nu\hbar) \tint_0^T[\beta^{-1} q(t)^2\,\dot{p}(t)^2+\beta \,q(t)^{-2}\,\dot{q}(t)^2]\;dt }\;\D(p)\,\D(q)\;.\en

Now, let us turn our attention to how coherent states, and the half-harmonic oscillator as well, have more to teach us about how certain path integrals can be constructed. 

\section{Canonical and Affine Coherent States}
\subsection{Canonical coherent states, along with their \\Fubini-Study metric}
The canonical coherent states employ the basic classical variables, $-\infty < p\;\&\:q<\infty$, and the quantum operators, $P\:\&\;Q$, for which $[Q,P]=i\hbar 1\!\!1$, and are given by
    \bn |p,q\> =e^{-iqP/\hbar}\,e^{ipQ/\hbar}\,|\omega\>\;, \label{p-1} \en
    where we choose  $(Q+iP/\om)|\om\>=0$, which implies that $\<\om|Q|\om\>=\<\om|P|\om\>=0$. 
    
    Two important relations are 
    \bn \<p',q'|p,q\>= e^{i(p'+p)(q'-q)/2\hbar - [
    \om^{-1} (p' - p)^2 +\om(q' - q)^2]/2\hbar}
   \;, \label{p-2} \en
   and $ \tint |p,q\>\<p,q| \; dp\,dq/2\pi\hbar = 1\!\!1 $, the identity operator.\footnote{A notable 
   story about canonical path integration is [9].}
    \subsubsection{The Fubini-Study metric for canonical quantization}
    An important expression of coherent states is found in the Fubini-Study metric [10], which is designed to be independent of any simple phase change the coherent states may have, i.e., $|p,q;f\>= e^{if(p,q)/\hbar}\,|p,q\>$, and is given by
    \bn d\sigma_{CQ}^2=2\hbar[\;|\!|\;d|p,q\>|\!|^2-|\<p,q|\;d|p,q\>|^2] = \omega^{-1}\,dp^2+\omega\,dq^2\;, \label{p- 3} \en
  and this leads to a flat surface with Cartesian variables also known as a `constant zero curvature'. This metric will have an important role to play in canonical path quantization.\footnote{It is noteworthy that, for a constant $c>0$, and  $p\ra p/c $ and $q\ra cq$, not only is the phase-space measure $dp\wedge dq$ invariant, but the CQ Cartesian metric $\omega^{-1}dp^2+\omega dq^2 $ remains Cartesian. }

    \subsection{Affine coherent states, along with their \\Fubini-Study metric}
    The affine coherent states employ the basic operators, which are $D\;\&\;Q$, provided, for this example, that $q>0\ra Q>0$. The affine coherent states are given (note: $p,q \ra p;q$) by
     \bn |p;q\>=e^{ipQ/\hbar}\,e^{-i\ln(q)D/\hbar}\,|\B\>\;, \label{p- 4} \en
     where $[(Q-1\!\!1)+iD/\B\hbar]\,|\B\>=0$, which implies that 
     $\<\B|Q|\B\>=1$ and $\<\B|D|\B\>=0$.

     The overlap of two affine coherent states is given [11], 
     and recalling that $q'\;\&\;q>0$, by
 \bn \<p';q'|p;q\>=\{ [(q'/q)^{1/2} +(q/q')^{1/2}]/2+i(q'q)^{1/2}(p'-p)/2\beta\hbar \,\}^{-2\beta} \;, \label{p- 6}\en
      and the resolution of identity is given by $ \tint|p;q\>\<p;q|\,\;[(1- 1/(2\beta)]dp\,dq/2\pi\hbar =1\!\!1$, provided $\beta>1/2$.
      
      The value of $\beta$ is determined   by the fiducial vector. In that case, it may be  occasionally reasonable to choose  $\beta=1$ to make things easier. In this  case, and still with $q'\;\&\;q>0$,
  \bn \<p';q'|p;q\>= \{[(q'/q)^{1/2} +(q/q')^{1/2}]/2+i(q'q)^{1/2}(p'-p)/2\hbar \,\}^{-2} \;, \ \label{p- 7}\en
      and $ \tint|p;q\>\<p;q|\,\; dp\,dq/4\pi\hbar =1\!\!1$, We mow restore $\beta$ again to see its role.
      \subsubsection{The Fubini-Study metric for affine quantization}
     The Fubini-Study metric [10, 11] for the affine coherent states is given by
     \bn d\sigma_{AQ}^2=2\hbar[\;|\!| \;d|p;q\>\,|\!|^2 -|\<p:q|\,d|p;q\>|^2]= (\B\hbar)^{-1}q^2\,dp^2+ (\B\hbar)\,q^{-2}\,dq^2\;,\ \label{p- 8}\en
     which is certainly not Cartesian, but is equally important because this metric is a `constant negative curvature' of magnitude $-2/\B\hbar$
    [12].\footnote{Observe that this new metric is {\it invariant} under the transformation  used in the previous footnote, i.e., when  a constant $c>0$ changed the variables,   $p\ra p/c $ and $q\ra cq$. That is the only rational metric to preserve its curvature under such a change.} 
    This metric will have an importantk role to play in affine path quantization.

      \subsubsection{An affine path integral of the half-harmonic oscillator}
 The classical Hamiltonian of the half-harmonic oscilkator
  is, as Eq.~(1) notes, still $H=(p^2+q^2)/2$, 
  but the important property is that $q>0$ and any path $q(t)$ just `bounces back off' a virtual wall at $q=0$. The affine path integration of this model is given by
  \bn &&\<p'';q'';T|p';q';0\> =\lim_{\nu\ra \infty  }\N_\nu\int_e^{(i/\hbar)\tint_0^T[p(t)\dot{q}(t)-(p(t)^2+q(t)^2)/2 ]\;dt}\no \\
   &&\hskip2em \times e^{-(1/2\nu\hbar)\tint_0^T[(\B\hbar)^{-1} q(t)^2\,\dot{p}(t)^2+(\B\hbar) \,q(t)^{-2}\,\dot{q}(t)^2]\;dt }\;\D(p)\,\D(q)\;.\ \label{p- 9}\en
   
   \subsubsection {A canonical version of an affine problem}  
   The following procedure  opens an interesting issue. Can a canonical approach using Eq,~(1) lead to the same results that the affine approach of Eq.~(13)? Let us change the classical variables in (13) to become
   $pq =r $ and $q=e^s$, in which $q>0$ while $-\infty<s<\infty$. 
   
   We also observe that while $[Q,P]=i\hbar1\!\!1$ for canonical quantization, it follows that $[e^Q,P]=i\hbar \,e^Q$ indicating that 
 the basic canonical  quantum variables can become the basic affine quantum variables in which $e^Q>0$. By introducing $P$ and $e^Q$, we can now see how to switch between CQ and AQ. In principle, that should permit us to  switch an affine quantization into a canonical quantization. %and thus vice versa! 
 
 In  the next section we try to do just that for the half-harmonic oscillator. To avoid confusion, we relabel $P\ra R$ and $Q\ra S$, for which $[S,R]=i\hbar1\!\!1$. and $[e^S,R]=i\hbar\,e^S$, with $e^S>0$. 
 In the previous line, for affine operators, it is evident that 
 $R$ plays the role  of $D$, and $e^S>0$ plays the role of $Q>0$. To obtain the Lie expression of affine quantization, the uniqueness of $e^S$ is ensured.
 
  \subsubsection{An AQ $\ra$ CQ path integration}
  %%%%%%%%-4
   We start with
   $q(t)\,dp(t)=dr(t)$ and $dq(t)/q(t)=ds(t)$, which makes $dp(t)\wedge dq(t)=dr(t)\wedge ds(t)$. This changes  $\D(p)\;\D(q) \ra \D(r)\;\D(s)$, and the path integration becomes
   \bn &&\hskip-2em \<r'',s'': T|r', s': 0\>= \lim_{\nu\ra \infty}\N_{\nu}\int e^{(i/\hbar) \tint_0^T \{ r(t)\,\dot{s}(t)-[r(t)^2/e^{2s(t)}+e^{2s(t)}]/2\hbar\}\;dt} \no \\
   &&\hskip6em \times e^{-(1/2\nu)\tint_0^T  [ (\B\hbar)^{-1} \dot{r}(t)^2+(\B\hbar)\,\dot{s}(t)^2] \;dt } \D(r)\,\D(s)\;.\label{p- 10}\en
   While Eq.~(1) contains an unusual semi-classical term in its Hamiltonian, the expression in Eq.~(5) offers yet another formulation that may not equate with that of (1).  Maybe we need to include some $\hbar$ terms already in the {\it semi}-classical nature of the usual classical action function because the tern $(i\hbar)$ signals that $\hbar$ has appeared, and especially when a term such as $3\hbar^2/4q^2$ is truly relevant, and $q$ is especially extremely  tiny,  it still preserves $q>0$. Even the collection of path integrated examples [1] allowed elements involving $\hbar$ in some of the `classical terms' in their path integral
   exponents. 
   
   We need a formulation that can
   {\it guaranty } validity!
  Fortunately, there is  still a formulation that primarily involves only coherent states, which also has positive features from both the simple Feynman approach and the complex Wiener type approach, which we now introduce.
 
 \section{Coherent State Canonical and Affine \\Ptoperties}
 \subsection{Coherent state canonical properties  for \\path integrals}
 
 In particular, the quantum Hamiltonian can be mapped into a semi-classical Hamiltonian by using suitable coherent states such as
  \bn |p,q\>=e^{-iqP/\hbar}\,e^{ipQ/\hbar}\,|\om\>\;,\ \label{p- 11}\en
 where the normalized fiducial vector $|\om\>$ obeys  $(Q+iP/\om)|\om\>=0$, and 
    an equation which connects the semi-classical and quantum Hamiltonians together is
     \bn  &&H(p,q) =\<p,q|\H(P,Q)|p,q\>=\<\om|\H(P+p,Q+q)|\om\>\no\\
      &&\hskip3.5em=\H(p,q)+ \mathO (\hbar;p,q) \;, \ \label{p- 12}\en
      where   $H(p,q) $ is a semi-classical term that may contain elements with an $\hbar $ dependence. When $\hbar\ra0$, then $H(p,q)=\H(p,q)$.

  \subsection{Coherent state affine properties for\\ path integrals}
  Generally, the quantum Hamiltonian can be mapped into a semi-classical Hamiltonian by using suitable coherent states, with $q>0$, such as
  \bn |p;q\>= e^{ipQ/\hbar }\,e^{-i\ln(q)\,D/\hbar}\,|\B\>\;,\ \label{p- 13}\en
 where the normalized fiducial vector $|\B\>$ obeys  $[(Q-1\!\!1)+iD/(\B\hbar)]|B\>=0$, and 
    an equation which connects the semi-classical and quantum Hamiltonians together is
     \bn  &&H'(pq,q) =\<p;q|\H'(D,Q)|p;q\>=\<\B|\H'(D+pqQ,qQ)|\B\>\no\\
      &&\hskip4.16em=\H'(pq,q)+ \mathO'(\hbar;pq,q) \;, \ \label{p- 14}\en
      where   $H'(pq,q) $ is a semi-classical term that may contain elements with an $\hbar $ dependence. Once again, if $\hbar\ra0$, then 
      $H'(pq,q)=\H'(pq,q)$.

 \section{Constraints in Path Integration}
Some examples of interest involve constraints and how to deal with them, and we need to briefly examined how path integration deals with them [7, 8, 13 - 17].

 The use of constraints reduces certain traditional physical aspects. It can generally be used  in path integration. For a simple, but risky, example, the equation
 \bn &&\N\int\!\int e^{(i/\hbar)\{\tint_0^T \{ p_1(t)\dot{q}_1(t)
+ p_2(t)\dot{q}_2(t)-\alpha(t)[p_2(t)^2+q_2(t)^2] \} } \no \\ &&\hskip3em \times e^{ -(i/\hbar)\{
 [p_1(t)^2 + p_2(t)^2] -V(q_1(t),q_2(t))\}\;dt}
 \no \\ &&\hskip 6em \times \;
\;\D(p_1,p_2)\;\D(q_1,q_2) \;\;\D(\alpha) \;, \label{p- 22}\en which introduces a $\Pi_{t=0}^T\;(\delta(p_2(t)^2+q_2(t)^2)$ by a properly normalized $\alpha(t)$ integration that leads to $p_2(t)^2+q_2(t)^2 =0$, i.e., $p_2(t)=0\;\&\;q_2(t)=0$ for all $0<t<T$. This equation relies on the integral where $\tint\!\tint \delta(x^2+y^2)\;dx\,dy/\pi =\tint\!\tint \delta(r^2) r\,dr\;d\theta/\pi=1$. However, if the term in question was $ [p_2(t)^2+q_2(t)^2)]^k$, and had any power other than $k=1$ it would cause considerable difficulty. Moreover, this approach effectually   overlooks the fact that $P_2^2+Q_2^2>0$, which then makes it a second-class constraint. It is important to observe that a re-linearized version of the combined constraints, i.e., $P_2^2+Q_2^2\ra aP_2^2+b\,Q_2^2$, where $0<a\;\&\;b<\infty$, only leads to different eigenvalues and a modest rescaling of the eigenfunctions. However, the number of eigenfunctions will remain the same, 
 hence the size of the Hilbert space  is not changed. For simplicity, we retain the initial expression, $P^2+Q^2$. 
     
     A safer procedure is to reduce the Hilbert space directly. For example, we introduce projection factors where $\E=\E^2=\E^\dag$, e.g., $ \E=\E(P^2+Q^2\leq\delta(\hbar)^2)$, which can deal with a second-class constraint, which is zero classically but non-zero when quantized. Such elements reduce a Hilbert space by accepting fewer vectors, such as  $\E |yes_j\>=|yes_j\>$ and eliminates other vectors, e.g., $\E|no_j\>=0$, even though $\<no_j|no_j\> >0$, $j=1,2,3...$. In this case, we could insist that $\<II|\E e^{-i\H T/\hbar} \E|I\>= \<II|\E e^{-i(\E \H\E) T/\hbar} \E|I\>$, which fits any Hilbert space reduction. 
     
     Most commonly, we can choose $\E =\sum_{l=1}^L \,|l\>\<l|$, with $\< l|l\>=1$ and $\<l'|l\>=0$ if $ |l'\>\neq|l\>$, which can limit some operators to a few , like $L<5$, distinct eigenvectors with the lowest eigenvalues. In such  cases, we could even choose $\E[(P_2^2+Q_2^2)^k \leq c_k\,
\hbar^k ]$, where $k>0$, which is a constraint that leads to  zero when $\hbar\ra0$. 

In different situations, one might  consider
 $\E(a<Q<b) =\tint_a^b |x\>\<x|\;dx$, where the limits do not cover the whole real line, and thus offer special projection operators, $\E$, that are not designed to  eliminate their results if $\hbar\ra0$. An earlier example of that kind is the half-harmonic oscillator for which $q>0$, and thus entails  $\E(Q>0)=\tint_0^\infty |x\>\<x|\;dx$.

 \subsection{Creating the $\E$ elements}
 A simple example will show how such terms can properly pass from the classical to the quantum realm. Consider
 our former  path integral example
  \bn &&\N\int\!\int e^{(i/\hbar)\{\tint_0^T \{ p_1(t)\dot{q}_1(t)
+ p_2(t)\dot{q}_2(t)-\alpha(t)[p_2(t)^2+q_2(t)^2] \} } \no \\ &&\hskip3em \times e^{ -(i/\hbar)\{
 [p_1(t)^2 + p_2(t)^2] -V(q_1(t),q_2(t))\}\;dt}
 \no \\ &&\hskip 6em \times \;
\;\D(p_1,p_2)\;\D(q_1,q_2) \;\;\D(R(\alpha))\;, \en
in which we have introduced $R(\alpha)$, where  $\alpha=\alpha(t)$. 

In a general sense, suppose   we first integrate the phase-space quantities, for a general problem, using  ${\bf T} $ for time-ordering, so that it leads to \bn && \<p'',q'':T|p',q':0\>\no \\ && \hskip2em
= \<p'',q''|\;{\bf T} e^{-(i/\hbar)\tint_0^T [\H_0(P,Q) + \alpha(t) \Phi(P,Q)]\; dt}
|p',q'\>\; \D R(\alpha)\; \label{p- 23}\en
in which $\Phi(P,Q)=P_2^2+Q_2^2\,$ refers to the constraints and $\alpha(t)$ is the Lagrange multiplier meant to enforce the constraint operator along with $\H_0(P,Q)$, which is the remainder of the original Hamiltonian operator, all of which now being integrated by the new measure, $\D(R(\alpha))$. For each instant of time, we introduce 
    \bn &&  \lim_{z \ra  0+} \lim_{L\ra \infty } \int^L_{-L} e^{-(i/\hbar)  \ep\,\alpha \,\Phi(P,Q)} 
    \sin[\ep(\delta^2 +z)\alpha/\hbar]/
    (\pi\alpha)  \; d\alpha \no \\ && \hskip6em
    =\E(\Phi(P,Q)\leq \delta(\hbar)^2)\;. \label{p- 24}\en

  \section{The Path Integration of Gravity}
 At last, we can tackle our principal topic. 
   
 We first recall the Arnowitt, Deser, and Misner version of the classical Hamiltonian [18], as originally expressed in the usual classical variables, 
 namely the momentum,  $\pi^{ab}(x)$, the metric, $g_{cd}(x)$, the metric determent 
  $g(x)=\det[g_{ab}(x)]$, and $^{(3)}\!R(x)$, which is the Ricci scalar for 3 spatial variables. Now the ADM classical Hamiltonian is essentially given by
\bn && H(\pi,g)=\tint \{ g(x)^{-1/2}[\pi^{ac}(x)g_{bc}(x)\pi^{bd}(x) g_{ad}(x) \no \\&&\hskip6em
-\half\,\pi^{ac}(x) g_{ac}(x)\pi^{bd}(x) g_{bd}(x)] \no \\
&&\hskip12em + g(x)^{1/2}\;^{(3)}\!R(x) \}\;d^3\!x\;. \label{p- 25}\en

\subsection{Introducing the favored classical variables}
The ingredients in providing a path integration of gravity include proper coherent states, the Funini-Study metric which turns out to be affine in nature, and affine Wiener-like measures are used for the quantizing of the classical Hamiltonian. While that 
effort is only part of the story, it is an important portion to ensure that the quantum Hamiltonian is a bonafide self-adjoint operator.
The remaining tasks will be covered later.

According to  the ADM classical Hamiltonian, it can also be expressed in affine variables, namely by introducing the 
`momentric', a name that is the combination  of {\it momen}tum  and me{\it tric}; this item is also called the ``dilation variable'') becoming  $\pi^a_b(x) \;(\equiv \pi^{ac}(x) \,g_{bc}(x))$, along with  the metric $g_{ab}(x)$. The essential physical requirement is that $g_{ab}(x)>0$, which means that $ds(x)^2=g_{ab}(x)\;dx^a\;dx^b>0$, provided that $\Sigma_a(dx^a)^2>0$. 

Now the classical Hamiltonian, expressed in affine classical variables, is given by
\bn && H\equiv \tint H(x)\;d^3\!x
 =\tint\{ g(x)^{-1/2}[\pi^a_b(x)\pi^b_a(x) -\half\,\pi^a_a(x)\pi^b_b(x)] \no \\
&&\hskip10em + g(x)^{1/2}\;^{(3)}\!R(x) \}\;d^3\!x\;. \label{p- 26}\en

\subsection{The gravity coherent states}
Based on the principal operators, $\hat{\pi}^a_b(x)=[\hat{\pi}^{ac}(x)^\dag\,\hat{g}_{bc}(x)+\hat{g}_{bc}(x)\,\hat{\pi}^{ac}(x)]/2$ and $\hat{g}_{ab}(x)>0$, these operators
 offer a closed set of commutation relations given by
 \bn   &&[\hp^a_b(x),\s \hp^c_d(y)]=i\s\half\,\hbar\,\delta^3(x,y)\s[\delta^a_d\s \hp^c_b(x)-\delta^c_b\s \hp^a_d(x)\s]\;,    \no \\
       &&\hskip-.10em[\hg_{ab}(x), \s \hp^c_d(y)]= i\s\half\,\hbar\,\delta^3(x,y)\s [\delta^c_a \,\hg_{bd}(x)+\delta^c_b \,\hg_{ad}(x)\s] \;, \\
       &&\hskip-.20em[\hg_{ab}(x),\s \hg_{cd}(y)] =0 \;. \no   \label{p-27}\en

We now choose the basic affine operators to build our coherent states for gravity, specifically
\bn |\pi;\eta\rangle=e^{(i/\hbar)\textstyle{\int}\pi^{ab}(x)\,\hat{g}_{ab}(x)\,d^3\!x}\;e^{-(i/\hbar)\textstyle{\int}\eta^a_b(x)\,\hat{\pi}^b_a(x)\,d^3\!x}\,|\beta\rangle\,\,[\,=|\pi;g\rangle]\;.\en 
Note: the last item in this equation is the new name of these vectors hereafter. 
 
A new fiducial vector, also named $| \B \> $ but now different, has been chosen now in  connection with the relation $[e^{\eta (x)}]_{ab}\equiv g_{ab}(x)>0$,  while  $-\infty<\{\eta(x)\}<\infty$, and 
which enters the coherent states as shown, using $|\B\>$ as the new fiducial vector that is affine-like, and obeys 
$[(\hat{g}_{ab}(x)-\delta_{ab}1\!\!1)+ i \hat{\pi}^c_d(x)/\B(x)\hbar]|\B\>=0$. It follows that
$\<\B|\hat{g}_{cd}(x)|\B\>=\delta_{cd}$ 
and $\<\B|\hat{\pi}^c_d(x)|\B\>=0$, which leads to the form given by \bn && \<\pi;g|\hat{g}_{ab}(x)|\pi;g\>=[e^{\eta(x)/2}]^c_a \;\<\B|\hat{g}_{cd}(x)|\B\>\;[e^{\eta(x)/2}]^d_b\no \\
&& \hskip7.5em
= [e^{\eta(x)}]_{ab}=g_{ab}(x)>0\;. \en    

In addition, we introduce the inner product of two gravity coherent states, which is given by
\bn &&\hskip-2em   \<\pi'';g''|\pi';g'\>=\exp\Big\{{-2\int}\beta(x)\,d^3\!x  \no\\ &&
\hskip-1em \times\ln\big\{
\det\{\frac{[ {g''}^{ab}(x)+{g'}^{ab}(x)]+i/(2 \,\beta(x)\, \hbar)[{\pi''}^{ab}(x)-{\pi'}^{ab}(x)]}{\det[{g''}^{ab}(x)]^{1/2}
\,\det[{g'}^{ab}(x)]^{1/2}} 
\big\} \Big\}. 
\;\label{p-33}\en

Finally, for some $C$, we find the  Fubini-Study gravity metric to be
\bn && d\sigma_g=C\hbar [\,|\!|\;d|\pi;g\rangle|\!|^2-|\langle\pi;g|\;d|\pi;g\rangle|^2]  \\ &&
\hskip2em =
\textstyle{\int}[\;\beta(x)\hbar)^{-1}\,(g_{ab}(x)\,d\pi^{ab}(x))^2 +(\beta(x)\hbar)\,(g^{ab}(x)\,dg_{ab}(x))^ 2\;]\;d^3\!x\;, \no \label{p- 30}\en
which is seen to imitate an affine metric and  will provide a genuine Wiener-like measure for a path integration. In no way could we transform this metric into a proper Cartesian form, as was done  for the half-harmonic oscillator. That is because there is no physically proper Cartesian metric for the variables $\pi^{ab}(x)$ and $g_{cd}(x)$.

\subsubsection{A special measure for the Lagrange multipliers}

To ensure a proper treatment of the operator constraints,  we choose a special measure of the Lagrange multipliers, $R(N_a,N)$, guided by the following procedures.

The first step is to unite the several classical constraints by using \bn && \tint\!\tint e^{i ( y^a H_a(x)+ y H(x)) }\,W(u, y^a, y, g^{ab}(x))\;\Pi_a dy^a\,dy\no \\
 &&\hskip7em =e^{-i u[H_a(x) g^{ab}(x)H_b(x) + H(x)^2] }\no\\  &&\hskip7em \equiv e^{-i u H_v^2(x)}
 \label{p- 31}\en 
with a suitable measure  $W$.

An elementary Fourier transformation\footnote{In mathematics, the following function being Fourier transformed is known as (a version of) rect(u) = 1 for $|u|\leq1$, and $0$ for $|u|>1$.} given by
$ M\tint_{-\delta^2}^{\delta^2}\,
e^{i\ep\tau\,uy} \; dy/2=\sin(u\ep\tau\delta^2)/u$, using a suitable $M$, 
 which then ensures that the inverse Fourier expression, where $\ep$ represents a tiny spatial  interval and $\tau$ represents a tiny time interval, as part of a fully regularized integration in space and time, and $u$ is another part of the Lagrange multipliers, $N_a(n\ep)$ and $N(n\ep)$, which leads to
  \bn &&\lim_{\zeta\ra0^+} \,\lim_{L\ra\infty}
  \int_{-L}^L e^{-iu\ep \tau\H_v^2(x)}\, \sin( u \ep\tau(\delta^2 +\zeta))/(u \pi)\;du  \no \\ &&\hskip5em
  = \E(\ep \tau \H_v(x)^2 \leq \ep\tau\delta^2) 
  %=  \E( \H_v(x)^2\leq \delta^2) \
  \no \\ &&\hskip5em 
  =\E(\H_v(x)^2\leq \delta^2) \;. \label{p- 32}\en

  This expression covers all self-adjoint operators, 
  and leads to a self-adjoint $\H_v=\tint\H_v(x)\;d^3\!x$.

  Bringing  together our  present tools lets us first offer a path integral for the gravity overlap of two coherent states, as given by
  \bn && \langle\pi'';g''|\pi';g'\rangle=\lim_{\nu\rightarrow \infty}{\cal{N}}_\nu\int\exp[-(i/\hbar)\textstyle{\int_0^T\!\int}[(g_{ab}\,\dot{\pi}^{ab})
  \,\,d^3\!x\,
dt]  \\ &&\hskip-4em
\times\exp\{-(1/2\nu\hbar)\textstyle{\int_0^T\!\int}[(\beta(x)\hbar)^{-1}\,(g_{ab}\,\dot{\pi}^{ab})^2 +(\beta(x)\hbar)\,(g^{ab}\dot{g}_{ab})^2 ]\,\,d^3\!x\,dt\} \no\\ &&
\hskip3em
\times\Pi_{x,t}\Pi_{a, b} \,d\pi^{ab}(x,t)\,dg_{ab}(x,t) \no \\
 &&\hskip-2em
  =\exp\Big\{{-2\int}\beta(x)\,d^3\!x  \no\\ &&
\hskip0em \times\ln\big\{
\det\{\frac{[ {g''}^{ab}(x)+{g'}^{ab}(x)]+i/(2 \,\beta(x)\, \hbar)[{\pi''}^{ab}(x)-{\pi'}^{ab}(x)]}{\det[{g''}^{ab}(x)]^{1/2}
\,\det[{g'}^{ab}(x)]^{1/2}} 
\big\} \Big\} \no
\;, \en
where the second equation indicates what such a path integration has been designed to acheive
for its goal.

 \section{The  Affine Gravity Path Integral}
 By adding all the necessary tools, and implicitly 
having examined a regularized integration version in order to effectively deal with suitable  constraint projection terms, we have choosen $\E\equiv \E (\H_v^2\leq\delta(\hbar)^2)$ for simplicity here, all of which leads us to
\bn &&\langle\pi'';g'':T| \E|
\pi';g':0\rangle =
\langle \pi'';g''|\E \,e^{-(i/\hbar)\E\,T}  \E\,| \pi';g'\rangle 
\no \\ &&\hskip2em
=\lim_{\nu\rightarrow \infty}{\cal{N'}}_\nu\int\exp\{-(i/\hbar)\textstyle{\int_0^T\!\int} [g_{ab}\,\dot{\pi}^{ab}\,+N^aH_a+NH]
\,d^3\!x\, dt\}\no \\ &&\hskip-2em
\times\exp\{-(1/2\nu\hbar)\textstyle{\int}_0^T\!\int[(\beta(x)\hbar)^{-1}\,(g_{ab}\,\dot{\pi}^{ab})^2 +(\beta(x)\hbar)\,(g^{ab}\,\dot{g}_{ab})^2]\,\,d^3\!x\,dt\}\no\\ &&\hskip5em 
\times[\Pi_{x,t}\Pi_{a, b} \,d\pi^{ab}(x,t)\,dg_{ab}(x,t)\,]\,{\cal{D}}R\{N^a, N\} \;, \label{p- 34}\en
in which the measure $R(N^a,N)$ is defined so that 
the operators,  $\H_a $ and $\H$, only support a sample of non-zero eigenvalues, e.g., $\E(\H_v^2 \leq \delta(\hbar)^2)$, where, e.g.,  $\delta(\hbar)^2\sim c\,\hbar^2$, or some other  tiny value that vanishes if $\hbar\ra0$. If $\H_v^2\leq\delta(\hbar)^2$ consists only of a continuous spectrum, a procedure to deal with that has been discussed in [15 - 20].

We let the reader choose their own regularization of the last equation to ensure that the $\ep$ terms are proper, and that the $\ep^2$ terms -- and higher $\ep^K, \;K>2$, terms as well -- lead to a proper continuum limit. In so doing, the overlap of two gravity coherent states, as shown above in (32), could  be particularly useful.

\section{Conclusions}
While a path integration of gravity using affine techniques appears to be valid, the author believes that  a canonical quantization, by path integration or by any other procedure, will only lead to invalid results for gravity. The author also believes that affine quantization should proudly  stand side-by-side with canonical quantization, which then enables each problem that is studied -- and, hopefully, correctly solved -- by first choosing  the proper set of quantization  tools. 

The author has also used standard Schr\"odinger procedures to study quantum gravity. This has involved using affine quantization along with a suitable Schr\"odinger representation and 
 formulating  a proper Schr\"odinger equation that relies on standard affine quantization procedures. Several such articles include such topics [21 - 27].   
 
 Some related articles: An early, wide reaching, article  also focussed on  quantizing gravity, which is [28]. A more mathematical paper, which focusses on  path integrations using only CQ, is [29]. Another early  paper  deals with a variety   of path integral issues, and has some interesting comments about affine quantizations [30]. 
\ve\ve

{\bf References}\ve\ve

[1] C. Grosche and F. Steiner, ``Classification of Solvable Feynman Path Integrals'';
arxiv: hep-th/9302053v1.\ve

     [2] A. Dynin, ``A Rigorous Path Integral Construction in Any Dimension''; arXiv:math/9802058. \ve
     
      [3]  P.A.M. Dirac, {\it The Principles of Quantum Mechanics}, (Claredon Press, Oxford, 1958), page 114, in a footnote. \ve
      
       [4] J. R. Kkauder, ``The Benefit of Affine Affine Quantization'', Journal of Hugh Energy Physics, Gravity and Cosmology 6, 175-185 (2020); doi:10.4236; jhepgc.2020.62014; arXiv:1912.08047.
     \ve
      
      [5] L. Gouba, ``Affine Quantization on the Half Line'', {Journal of High Energy Physics, Gravitation and Cosmology} {\bf 7}, 352-365 (2021); 
 arXiv:2005.08696.\ve

 [6] C. Handy, ``Affine Quantization of the Harmonic Oscillator on the Semi-bounded 
Domain $(-b,\infty)$ for $b:0\ra\infty$''; arXiv:2111:10700. \ve

 [7] J. R. Klauder and I. Daubechies, ``Quantum Mechanical Path Integrals with Wiener Measures for all Polynomial Hamiltonians'', Phys.
 Rev. Letters {\bf 5}, 1161-1164 (1984).\ve

      [8] I. Daubechies and J. R. Klauder, ``Quantum-Mechanical Path Integrals with Wiener Measure for all
Polynomial Hamiltonians II'', J. Math. Phys. {\bf 26}, 2239-2256 (1985). \ve
     
     [9] J. R. Klauder, ``Coherent  States in Action'', 
 arXiv:quant-ph/9710 029v1. \ve
 
 [10] Wikipedia: ``Fubini-Study metric''.\ve
 
 [11] J. R. Klauder, ``On the Role of Coherent States in Quantum Foundations''; arXiv:1008.4307v1.\ve
 
 [12] Scholarpedia: ``negative curvature 2d''.\ve

  [13] D.M. Gitman and I.V. Tyutin, {\it Quantization of Fields with Constraints}, 
(Springer-Verlag, Berlin, 1990).\ve

[14] J. R. Klauder, ``Coherent State Quantization of Constraint Systems'', Ann. Phys. {\bf 254}, 419 (1997); arXiv:9604033v1.\ve

[15] A. Kempf and J. R. Klauder, ``On the Implementation of Constraints through Projection Operators''; arXiv:quant-ph/0009072v1.\ve

  [16] J. R. Klauder and S. V. Shabanov, ``An Introduction to Coordinate-free Quantization and its Application to Constrained Systems''; 
  arXiv:quant-ph/9804049v1. \ve

[17] W. R.  Bomstad and J. R. Klauder, ``Linearized Quantum Gravity Using the Projection Operator Formalism"; arXiv:gr-qc/0601087v2. \ve

[18] R. Arnowitt, S. Deser, and C. Misner, in {\it Gravitation: An Introduction to 
Current Research}, Ed. L. Witten, (Wiley \& Sons, New York, 1962), p. 227; arXiv:gr-qc/0405109.\ve

[19] J. R. Klauder, {\it A Modern Approach to Functional Integration}, \\(Bitkh\"auser, 2011). \ve

       [20] J. R. Klauder, ``Universal Procedure for Enforcing Quantum Constraints'', Nuclear Physics B {\bf 547}, 397-412 (1999);
     https://www.sciencedirect.
     com/journal/nuclear-physics-b. \ve
 
       [21] J. R. Klauder, ``Quantization of Constrained Systems'',  Lect. Notes Phys. {\bf 572}, 143-182 (2001); arXiv:hep-th/0003297v1. \ve

     [22] J. R. Klauder,  ``Quantum Gravity Made Easy'', {Journal of High Energy PhysicsGravitation and Cosmology} {\bf 6}, 90-102 (2020); doi: 10.4236/
     jhepgc.2020.61009.\ve

     [ 23] J. R. Klauder, ``Building a Genuine Quantum Gravity'', {Journal of High Energy Physics, Gravitation and Cosmology} {\bf 6}, 159-173 (2020); \\doi:10.4236/jhepgc.2020.61013.\ve

  [24] J. R. Klauder,   ``Quantum Gravity, Constant Negative Curvatures, and Black Holes'', {Journal of High Energy Physics, Gravitation and Cosmology} {\bf 6}, 313-320 (2020); doi:10.4236/jhepgc.2020.63024.\ve
     
   [25] J. R. Klauder,  ``Using Affine Quantization to Analyze Non-Renormalizable Scalar Fields and the Quantization of Einstein's Gravity'', {Journal of High Energy Physics, Gravitation and Cosmology} {\bf 6}, 802-816 (2020); doi:10.4236/jhepgc.2020.64053.\ve
     
   [26 ] J. R. Klauder,  ``The Unification of Classical and Quantum Gravity'', {Journal of High Energy Physics, Gravitation and Cosmology} {\bf 7}, 88-97 (2021); doi:10.4236/jhepgc.2021.71004.\ve
     
  [27] J. R. Klauder,   ``Using Coherent States to Make Physically Correct Classical-to-Quantum Procedures That Help Resolve Nonrenomalizable Fields Including Einstein’s Gravity'', {Journal of High Energy Physics, Gravitation and Cosmology}  {\bf  7}, 1019-1026 (2021); doi:10.4236/6
  jhepgc.2021.73060.\ve
  
  [28] J. R. Klauder, ``Noncanonical Quantization of Gravity. I. Foundations of Affine Quantum Gravity''; arXiv:gr-qc/9906013v2, \\ve

[29] J. R. Klauder, ``Essential aspects of Wiener-measure Regularization for Quantum Mechanical Path Integrals'', Nonlinear Analysis {\bf 63} e1253 – e1261 (2005).\ve

  [30] L. Hartmann and J. R. Klauder, ``Weak Coherent State Path Integrals'', J. Math. Phys. {\bf 45}, 87 (2004); https://doi.org/10.1063/1.1627959.

\end{document}